\documentclass[preprint,authoryear,12pt,twocolumn]{elsarticle}

\usepackage{epsfig}
\usepackage{amssymb}
\usepackage{aas_macros}

\usepackage[ps2pdf,%
a4paper=true,%
breaklinks=true,%
colorlinks=true,%
pdfauthor={First Author et al.},%
pdftitle={Template for manuscripts in Advances in Space Research}%
]{hyperref}

\journal{Journal of Quantitative Spectroscopy and Radiative Transfer}
\biboptions{authoryear}

\newcommand{\mdot}{\mbox{$\dot{M}$}}
\newcommand{\Rstar}{\mbox{$R_\ast$}}

\newcommand{\Teff}{\mbox{$T_{\rm eff}$}}
\newcommand{\Lbol}{\mbox{$L_{\rm bol}$}}

\newcommand{\myr}{\mbox{$M_\odot\,{\rm yr}^{-1}$}}

\newcommand{\msim}{\raisebox{-.4ex}{$\stackrel{>}{\scriptstyle \sim}$}}

\def\changed{}

\begin{document}

\begin{frontmatter}



\title{Moving inhomogeneous envelopes of stars}


\author{Lidia M. Oskinova}
\address{Institute for Physics and Astronomy, University of Potsdam,
14476 Potsdam, Germany}
\ead{lida@astro.physik.uni-potsdam.de}
\author{Brankica\ Kub\'atov\'a}
\address{Astronomick\'y \'ustav AV \v{C}R, Fri\v{c}ova 298, 251 65
Ond\v{r}ejov, Czech Republic \\
Matemati\v{c}ki institut SANU, Kneza Mikhaila 36, 110 01
Beograd, Serbia}
\author{Wolf-Rainer Hamann}
\address{Institute for Physics and Astronomy, University of Potsdam,
14476 Potsdam, Germany}


%

\end{frontmatter}

\noindent
{\em Abstract}: 
Massive stars are extremely luminous and drive strong winds, blowing a large 
part of their matter into the galactic environment before they finally explode 
as a supernova. Quantitative knowledge of massive star feedback is
required to understand our Universe as we see it. Traditionally, massive stars 
have been studied under the assumption that their winds are homogeneous and 
stationary, largely  relying on the Sobolev approximation. However, 
observations 
with the newest instruments, together with progress in model calculations, 
ultimately dictate a cardinal change of this paradigm: stellar winds are highly
inhomogeneous. Hence, we are now advancing to a new stage in our 
understanding of
stellar winds. Using the foundations laid by V.V. Sobolev and his school,
we now update and further develop the stellar spectral analysis techniques. 
New sophisticated 3-D models of radiation transfer in inhomogeneous 
expanding media elucidate  the physics  of stellar winds and improve classical 
empiric mass-loss rate diagnostics. Applications of these new techniques 
to multiwavelength observations of massive stars yield consistent and robust  
stellar wind parameters.


\parindent=0.5 cm

\section{Introduction}

The initial mass of a star on the zero-age main sequence largely determines its 
fate. Stars born with masses exceeding $\sim 10\,M_\odot$ end their 
lives in a core-collapse event, e.g. a supernova (SN) explosion, and leave a 
neutron star or a black hole as remnant \citep{Heger2003}. Such   
massive stars are luminous, with bolometric luminosities exceeding $L_{\rm 
bol}\msim 10^4\,L_\odot$. On the main sequence, massive stars have 
spectral types earlier than B2V. These bright stars live very fast, the most 
massive of them die within just $\sim 10$\,Myr. Albeit we see many massive 
stars by naked eye in the night sky (e.g.\ the Orion Belt consists of massive 
stars), these stars are actually very rare and constitute 
only $\sim 0.4$\%\ of all stars in our Milky Way.   

Despite their small number, massive stars have enormous impact on the galactic 
ecology. Their strong ionizing radiation and stellar winds, as well as 
their final demise in  SN explosions, largely determine the physical 
conditions in the interstellar medium (ISM) and influence the formation 
of new generations of stars and planets. Thus, massive stars are among the 
key players in the cosmic evolution. 

The atmospheres of hot massive stars are usually transparent in the continuum 
but opaque in many spectral lines. Because the stars are hot, a large fraction 
of their bolometric luminosity is emitted at  ultraviolet (UV) wavelengths. 
The radiation leaves the star in radial direction. A photon in a spectral 
line $\nu_0$ may be absorbed by an ion and re-emitted in any direction, 
transferring its  momentum to the ion. The ion would accelerated. Because of 
the Doppler effect, the wavelength of the spectral line will shift, and will be 
able to scatter light with wavelengths other than $\nu_0$. Hence, the photons 
within a broad wavelength range  will be ``swept 
up'', by the same spectral line.  The Coulomb coupling between particles ensures 
the collective motion, and a stellar wind develops. Such radiatively driven 
stellar winds  \citep[][CAK]{CAK1975}  are ubiquitous in hot non-degenerate 
stars.

The amount of mass removed from the star by its wind is determined by 
the mass-loss rate, $\dot{M}$. Theory predicts that for 
O-stars the mass-loss rates are in the range $\dot{M}_{\rm CAK}\approx 
(10^{-7}-10^{-5})$\myr\ depending on the fundamental stellar parameters \Teff, 
\Lbol, and $\log{g}$ \citep{Pau1986, Vink2001}. Hence, during stellar life time, 
a significant fraction of mass is removed by the stellar wind. Thus, 
the mass-loss rate is a crucial factor of stellar evolution.

\section{Empirical diagnostics of mass-loss}

To check and validate theoretical predictions, robust empirical estimators
of mass-loss rates shall be employed. These diagnostics usually rely on 
a spectroscopic analysis. Below we briefly consider some common examples of 
such analyses. 

\subsection{Resonance lines}

For hot stars, the resonance lines of most important  ions are located in the 
UV part of the 
electromagnetic spectrum. When formed in a wind, these lines show P Cygni-type 
profiles \citep[see e.g.][]{LC1999}. 

The resonance line of an ion  is produced by photon scattering, 
therefore the line strength is a linear function of the density, 
which is related to the the expansion velocity $v(r)$ by the continuity 
equation 
\begin{equation}
 \rho(r)=\frac{\dot{M}}{4\pi\,v(r)\,r^2}.
 \label{eq:rho}
\end{equation}
The radial dependence of the wind velocity is usually prescribed by the 
``$\beta$-velocity law'', $v(r)=v_\infty (1-1/r)^\beta$. 

The line strength, the terminal wind velocity, $v_\infty$, and the parameter 
$\beta$ can be measured from the observed spectral
line. Hence, in principle,  by modeling a resonance line of an ion, 
the product of its ionization fraction and mass-loss rate  could be 
empirically obtained. To model a spectral line,  an adequate theory of line 
formation is required. 

Line formation in a moving stellar envelope was studied by 
V. V. Sobolev  \citep{Sobolev1960}. It was shown that if the 
thermal motions 
in the atmosphere can be neglected compared to the macroscopic velocity,  
the radiative transfer problem can be significantly simplified \citep[see 
review by][]{Grinin2001}. This is now known as the {\em Sobolev 
approximation}.  

The Sobolev approximation is well justified in stellar winds, and was 
extensively
used for their analysis. At least two different solution techniques that relied 
on the Sobolev approximation were developed  \citep{Castor1970,Lucy1971}. An 
atlas of theoretical P Cygni profiles was computed \citep{Castor1979} and used 
to estimate mass-loss rates from the first available UV spectra of O stars 
\citep[e.g.][]{Conti1980}. 

With time the limitations of the Sobolev approximation became clear. For 
instance, 
within a Doppler-shifts of a few tens of kilometers per second around the line 
center, 
the profiles computed in Sobolev approximation are inaccurate,
especially because of the high turbulence present in stellar winds 
\citep{wrh1981} and/or non-monotonic wind velocities \citep{Lucya1982,Lucy1983}.

These shortcomings were overcome by \citet{wrh1981}, who compared line 
profiles computed with a comoving frame approach \citep{Mihalas1975,Hamann1980} 
with those computed using Sobolev approximation. It was shown that the error in 
the Sobolev approximation arises mainly from the treatment of the formal 
integral and, to a lesser extent, from the approximated source function. Based 
on this suggestions \citet{Lamers1987} developed the ``Sobolev with Exact 
Integration'' method (SEI).  In this method, the source function is calculated 
in the Sobolev approximation, but the equation of transfer is 
integrated exactly. As a result, the model  provides significantly better fits 
to the observed lines \citep{Gro1989}, consequently allowing for more precise  
mass-loss rate determinations \citep{Lamers1999}. 

While UV resonance lines provide excellent mass-loss rate diagnostic, 
there are also serious difficulties. First of all, to measure UV spectra one 
needs a space-based observatory. Presently, UV spectroscopy is offered 
only by the {\em Hubble Space Telescope}, a highly oversubscribed instrument.  
An even more serious problem is that in
Galactic O type stars the resonance lines of the CNO elements are 
usually saturated. 
Therefore these lines are not sensitive to the precise values of the mass-loss 
rate and, hence, not suitable for mass-loss determinations.

\subsection{Stellar atmosphere model PoWR}
\label{sec:powr}

As an alternative to the SEI method, detailed non-LTE stellar 
atmosphere models for expanding atmospheres were developed in the last decades 
\citep[see review by][]{puls2008}. Such models do not 
rely on the Sobolev approximation but solve numerically the radiative 
transfer in the co-moving frame. The coupling between radiation field and 
statistical equations  is included, leading to a high-dimensional set of 
non-linear equations fully coupled in space and frequency. The synthetic 
emergent spectrum is calculated over a broad energy range. Comparing 
model and observed spectra allows to estimate stellar as well as wind 
parameters, specifically $\dot{M}$.   

An example of such advanced stellar atmosphere model is the PoWR code 
\citep[e.g.][]{Graf2002, Hamann2003, Sander2015}. It solves the non-LTE 
radiative 
transfer in a spherically expanding atmosphere simultaneously with the 
statistical equilibrium equations and accounts at the same time for energy 
conservation.  Complex model atoms with hundreds of levels and thousands of 
transitions are taken into account. The extensive inclusion of the iron group 
elements is important not only because of their blanketing effect on the 
atmospheric structure, but also because the diagnostic wind lines in the UV 
(e.g.\ the C\,{\sc iv} and Si\,{\sc iv} resonance lines) are heavily blended 
with the ``iron forest''.  X-ray emission and its effects on the ionization 
structure of the wind are also included in the model.  

PoWR models are extensively used for the spectral analysis of stars with 
strong winds, such as Wolf-Rayet (WR) stars across the broad range of 
metallicities \citep{Sander2012, Hainich2014, Hainich2015}. Because the 
hydrostatic and wind parts of the atmosphere are solved consistently 
\citep{Sander2015}, the PoWR models are well suited for modeling O star spectra 
\citep{Evans2011,Shenar2015}, as well as  B star spectra \citep{osk2011}.
Overall, PoWR models can be applied for spectroscopic analysis of any type of 
hot stars \citep[e.g.][]{Todt2010,Jef2010}.

\subsection{Optically thin recombination lines: H$\alpha$}

While the UV observations providing access to the resonance lines are 
scarce, the optical spectra of bright massive stars are easy to 
obtain. In OB supergiants, the H$\alpha$ line is typically in 
emission. This line provides a convenient diagnostic of the wind. 

In most O stars, the H$\alpha$ line is optically thin. In this 
case, the line luminosity is the volume integral over the line 
emissivity,  $j_{\rm l}$,
\begin{equation}
 L_{\rm l}=\int_V j_{\rm l}{\rm d}V.
 \label{eq:line}
\end{equation}

The H$\alpha$ line is mainly fed via the recombination cascade, i.e.\ an 
interaction between an electron and an ion. This is a two-body process 
that depends on the square of the density and is some function of the 
temperature, $j_{\rm l}\sim \langle\rho^2 \rangle f(T)$, where $\langle\rho^2 
\rangle$ 
is the average over the volume.

A medium with density fluctuations can be described using the wind 
inhomogeneity parameter \citep{allen1973}
\begin{equation}
{\cal X} \equiv \frac{\langle \rho^2 \rangle}{\langle \rho\rangle^2}.
\label{eq:x}
\end{equation}

Using the continuity equation Eq.\,(\ref{eq:rho}) for the averaged density 
$\langle \rho \rangle$, and combining Eqs.\,(\ref{eq:line},\ref{eq:x}), the 
mass-loss rate can be expressed as 
\begin{equation}
 \dot{M}\propto v_\infty \sqrt{\frac{L_{\rm l}}{\cal X}},
 \label{eq:xi}
\end{equation}
where $v_\infty$ and $L_{\rm l}$ can be measured from observed spectra,  
but the parameter ${\cal X}$ is largely unknown. 

The value of ${\cal X}$ is very difficult to estimate; moreover, it can vary 
from object to object. As a first approximation, it seem sensible to set
${\cal X}=1$, i.e. to assume that stellar winds are smooth on average. In this 
case, measuring the easily observable H$\alpha$ emission line can provide the 
mass-loss rate. 

\citet{puls1996} refined the theory of 
H$\alpha$ formation, and presented scaling relations that 
connect $\dot{M}$, $v_\infty$, stellar parameters, and the H$\alpha$ 
equivalent width. They applied this method to samples of O-stars in 
the Galaxy, the Large and the Small Magellanic Cloud. The results revealed  a 
tight metallicity-dependent relation  between the ``radius modified 
stellar wind momentum rate'', $\dot{M}v_\infty\sqrt{R_\ast}$, and the stellar 
luminosity. It was shown that H$\alpha$ based mass-loss rates of O stars are in 
good 
general agreement with those derived from radio measurements (being 
free-free emission, the latter also 
depend on $\langle\rho^2\rangle$ processes). Moreover, the H$\alpha$ based 
mass-loss 
rates of O stars were found in generally good agreement with  
theoretical predictions, i.e.\ in the range $10^{-4}\,...\,10^{-7}$\,\myr\ 
depending on spectral type.


\section{Stellar wind clumping}

Despite of this large progress in measuring mass-loss rates,  one has 
to be aware that the  ${\cal X}=1$ assumption is not really justified, 
on the contrary there is  strong evidence for stellar wind clumping 
\citep{Hamann2008}. 

Clear evidence of wind inhomogeneity was provided by the detection of
stochastic variability in  the He\,{\sc ii}\,$\lambda$4686\,\AA\ emission
line in the spectrum of an O supergiant \citep{Eversberg1998}, explained 
by  clump propagation. Line-profile variability of  H$\alpha$ was seen 
in a large sample of O-type supergiants, and attributed to the presence of 
shell fragments in structured winds \citep{Markova2005}. \citet{Prinja2010} 
demonstrated that the winds of B supergiants are clumped by using spectral 
diagnostics. In a recent study, \citet{Martins2015} showed that spectral lines 
of OB  supergiants are variable on various time scales likely because of the 
wind structuring. The  line-profile variations in a 
sample of WR and O stars were monitored by \citet{Lepine1999, Lepine2008}.  
The observations were explained using a phenomenological model  that depicts 
winds as being made up of a large number of randomly  distributed, radially 
propagating clumps.  

High-mass X-ray binaries (HMXBs) provide a confirmation of strong wind 
clumping. In some of these systems, a neutron star (NS) is in a close orbit 
deeply inside the stellar wind of an OB star. Accretion from the clumped 
stellar wind onto the NS powers strongly variable X-ray emission 
\citep[e.g.][]{osk2012,Martinez2014,Bozzo2016}. 
\citet{vdMeer2005} studied the X-ray light curve and spectra  of 4U
1700-37 and concluded that the feeding of the NS by a strongly clumped 
stellar wind is consistent with the observed stochastic variability. 
Further evidence of donor wind clumping comes from the 
analysis of the X-ray spectra.  \citet{Schulz2002} and \citet{Gim2015}
reviewed the  spectroscopic results obtained with X-ray  observatories
for wind-fed HMXBs.  They explained the observed spectra 
as originating in  a clumped stellar wind, where cool dense clumps 
are embedded in rarefied X-ray photoionized gas. Fluorescence lines 
were used to trace  wind clumps that are located close to the photosphere 
in a case of B-type supergiant donor \citep{Tor2015}. 

The first X-ray spectra of putatively single O-type stars were obtained 
with the 
{\em Einstein} X-ray observatory.  \citet{StewartFabian1981} 
used  {\em Einstein} spectra of the O supergiant $\zeta$\,Pup 
to determine its mass-loss rate. They applied a photoionization code and 
computed  the stellar wind  opacity using \mdot\ as a model parameter. The 
X-ray based mass-loss rate was lower by a factor of a few than 
obtained  from   H$\alpha$  and radio emission. As the most plausible 
explanation for this discrepancy it was suggested that the mass-loss rate 
measured from H$\alpha$ and radio emission is overestimated because of wind 
clumping. 

The {\em Rosat} X-ray spectrum of  $\zeta$\,Pup was investigated by 
\citet{hil1993}. The wind was assumed to be smooth with a mass-loss 
rate $\dot{M}=5\times 10^{-6}$\,\myr\, consistent with the H$\alpha$ 
measurement. It was found that the high opacity of the stellar  wind should 
completely block soft X-rays ($<0.5$\,keV).  However, since such  soft 
X-rays are observed, it was concluded that a significant fraction of the 
X-ray emitting
plasma is located far out in the wind, at distances $> 100$\,\Rstar. 

The launch of {\em XMM-Newton} and {\em Chandra} X-ray telescopes made 
high-resolution X-ray spectroscopy possible. The X-ray spectroscopic 
diagnostics allows to constrain the location of hot plasma emitting X-rays, and 
to probe the wind opacity \citep{Macfar1991,porq2001}. The analysis of 
X-ray spectra unambiguously showed that hot plasma is located not only far 
out in the wind, but also close to the stellar photosphere at distances $< 
1.5$\,\Rstar, and that the  wind opacity is much lower than expected from the 
H$\alpha$ derived mass-loss rates \citep[e.g.][]{wc2001,Kahn2001,cas2001}.  

Among others, the following two ways were suggested to explain the X-ray data: 
stellar wind 
clumping reduces the wind opacity and allows radiation to escape even for high 
mass-loss rates  \citep{feld2003,osk2004}, and stellar mass-loss rates are much 
lower than found from $\rho^2$-based diagnostics \citep{Kramer2003}.  

\section{P\,{\sc v} and the discordance of mass-loss rates diagnostics} 

\citet{massa2003} studied Far Ultraviolet Spectroscopic Explorer ({\em FUSE}) 
spectra of O stars in the Large Magellanic Cloud using the SEI method. They 
highlighted the importance of the P\,{\sc v} $\lambda\lambda 1117, 
1128$\,\AA\ resonance doublet for mass-loss  diagnostics. This resonance 
doublet is never saturated because of the low phosphorus abundance  
(e.g.\ 1000 times less than carbon) making it specially suitable for 
mass-loss diagnostics. Moreover, P\,{\sc v} is the dominant ionization 
stage in O stars, hence its ionization fraction is nearly unity 
\citep{Krt2012}. However, 
the analysis of observed spectra by \citet{massa2003} revealed much weaker lines 
of (P\,{\sc v}) than expected if the H$\alpha$ based mass-loss rates were true. 
It was concluded that the weak P{\sc v} lines imply that either $\dot{M}$ is 
very low, or the assumed abundance of phosphorus is too large, or the winds are 
strongly clumped. 

\citet{Bouret2005} and \citet{Fullerton2006}  demonstrated the 
severe discordance of mass-loss rates empirically obtained from the  
$\rho^2$- (such as H$\alpha$ and radio)  and the $\rho$-based (such as 
resonance lines, e.g. P\,{\sc v}) diagnostic methods.
The mass-loss rates measured from the UV lines were found to be at least an 
order 
of magnitude lower than those measures from H$\alpha$. The UV line   
diagnostics were considered more reliable than H$\alpha$, 
because they do not depend on clumping.   

To explain this discordance it was suggested that the clumping is  very strong 
and therefore mass-loss rates measured from H$\alpha$ have to be reduced by 
orders of magnitude. Thus, based on the UV and X-ray diagnostics, the 
empirically estimated 
mass-loss from O-type stars were found to be significantly lower that 
predicted by the theory. This severe reduction of mass-loss rates was the 
basis for 
\citet{so2006} to  propose a new evolutionary scenario according to which the 
stars with initial mass above $40\,...\,50\,M_\odot$ loose the bulk of their 
mass not via radiatively driven stellar winds but via explosive eruptions, and 
to discuss the possibility that Wolf-Rayet stars may not be the descendents 
of such massive stars. 

\section{Microclumping}

Clearly, it is necessary to re-evaluate how clumping is accounted 
for in the spectral analysis and re-investigate the assumptions on which 
the mass-loss diagnostics are based. The usual and but stringent 
approximation is that all clumps in stellar wind are optically thin. 
This approximation, called  {\em microclumping}  or  filling factor 
approach, was implemented in non-LTE codes already back in 1990s
\citep{hk1998,Hil1999}.
 
If we assume that  the density is uniform inside the clumps while the 
interclump medium is void, then ${\cal X}=D$, where $D$ is the density 
enhancement  
within clumps as compared to a smooth model with same mass-loss rate \mdot. In 
this case, the volume filling factor of the clumps is $f_{\rm V} = D^{-1}$. In 
the models, the rate equations have to be solved only for the clump medium, 
where the density is $D\rho$ instead of $\rho$ as in the smooth case.

Consequently, in the radiative transfer equation, the smooth-wind opacity and
emissivity $\kappa(\rho)$ and $\eta(\rho)$ must be replaced for a
clumped wind by 
\begin{equation}
\kappa_{\rm f} = f_{\rm V}\ \kappa_{\rm C}(D \rho) ~~~\mathrm{and}~~~ 
\eta_{\rm f} = f_{\rm V}\ \eta_{\rm C}(D \rho)\,
\label{eq:kappa_D}
\end{equation}
where $\kappa_{\rm C}$ and $\eta_{\rm C}$ are the non-LTE opacity and emissivity 
of the clump matter.
The atomic transitions that contribute to the opacity and emissivity
scale with different powers of the density.  For processes linear in density, 
$f_{\rm V}$ and $D$ cancel. However, empirical mass-loss diagnostics 
are often based on processes that scale with the square of the density 
(recombination lines, free-free  emission). 
When the wind is clumped, the emitted  flux is enhanced by a factor of 
$D$ compared to a homogeneous model with the same mass-loss rate. Consequently, 
when a given (free-free radio or or recombination-powered line) emission is 
analyzed with a model that accounts for microclumping, the derived mass-loss 
rate will be lower by a factor of $\sqrt{D}$ than obtained with a smooth-wind 
model (see Eq.\,\ref{eq:xi}). 

In WR star spectra, fitting the observed electron scattering wings of 
strong emission lines  can be used  to determine the clumping
factor $D$ and  its radial dependence \citep{hil1991,hk1998}. For WR stars
this method yields  typical clumping factors $D$ between 4 and 10. 

Unfortunately, this method is not  applicable for O stars, since their spectra 
do not show suitably strong  emission lines. Hence, $D$ has to be constrained
indirectly, e.g. as as a model parameter. Often, a radial dependence of 
clumping parameter is allowed in the models. \citet{Puls2006} estimated that 
the clumping factor $D$ is about four times larger in the 
line-forming region, compared to the radio-emitting region far away from the 
star.  From their multiwavelength analysis of a B supergiant spectrum
\citet{Puebla2016} estimated that $f_\infty$, i.e.\ the smallest filling 
factor approached at large wind velocities ($v(r)\rightarrow v_\infty$),  
is 0.01, while \citet{Bouret2012} derived $f_\infty=0.03\,..\,0.06$ for their 
sample of O stars. These results imply that the derived mass-loss rates are 
reduced by a factor 4\,...\,10 compared to the smooth wind analyses.

As an interesting consequence,  microclumping  reducing the size of the 
photoionized region in HMXBs. Even the largest clumps in stellar wind are 
likely optically thin at hard X-ray wavelengths emitted by an accreting NS.
The strong X-ray radiation photoionizes the  wind region surrounding the NS 
\citep{Hat1977}, and affects wind driving \citep[e.g.][]{Krt2015}.  
In clumped stellar winds the recombinations are favored compared to the 
ionizations. Hence,  the size of the area photoionized by X-rays 
is smaller in realistic clumped winds compared to the smooth wind case 
\citep{osk2012}.

\section{Macroclumping}
\label{sec:osk07}
While the microclumping approximation is very convenient, this approach 
is too stringent. In reality stellar winds clumps could be optically 
thick at some frequencies. This situation has to be accounted for
in modeling, e.g.\ using the {\em macroclumping} approach.

Accounting for optically thick clumping is required in a large variety 
of problems, such as continuum driven winds in luminous blue variables (LBV),  
X-ray emission from massive stars, or resonance line formation in stellar 
winds. 

\subsection{Grey opacity}

\citet{Shaviv1998} and \citet{Shaviv2000} considered the gray opacity in a 
stellar 
atmosphere, where the photon mean free path does not exceed the scale of 
inhomogeneities. In such multi-phase porous atmosphere the radiation is able 
to escape easier while exerting a weaker average force. For a continuum driven 
wind, 
a star with the same luminosity would experience considerably smaller mass-loss 
compared to the case of homogeneous atmosphere. This may explain the 
observations of the LBV star $\eta$\,Car \citep{Shaviv2000}. 
\citet{Owocki2004} introduced a ``porosity-length'' formalism to derive a simple 
scaling for the reduced effective opacity and used this to obtain an associated 
scaling for the continuum-driven mass-loss rate from stars that formally exceed 
the Eddington limit. 
\citet{Quataert2016} developed analytical and numerical models of the 
properties of super-Eddington stellar winds. \citet{Brown2004} pointed out that 
optically-thick clumping leads to the reduction of multiple scattering 
and, consequently, photon momentum delivery.

\subsection{Continuum opacity}

Stellar winds of OB stars are effective in absorbing X-rays via  
bound-free and K-shell photoionization processes. This high continuum 
opacity is strongly wavelength dependent \citep[e.g.][]{ver1995}.   
In inner wind regions (e.g.\ at 0.5$v_\infty$) of a typical O supergiant, 
the mean free path of a photon at soft X-ray wavelength (e.g.\ at 
19\,\AA) is only $\sim 10^{-2}\,\Rstar$. Clumps may have larger sizes and, 
hence, be optically thick. 

{\changed The geometrical shape of clumps is not observationally constrained. 
There is only limited evidence for an intrinsic polarization of O star 
 \citep{McDavid2000,Harries2002}, albeit this may be theoretically 
expected for stars with stronger winds and strong clumping 
\citep{Brown2000,Li2000,Davies2007}. Sophisticated but 1-D hydrodynamic 
models predict that wind clumps are flattened structures, with only small 
extend in radial direction \citep{Owocki1988,feld1997}. An initial attempts to 
include the multi-dimensional nature of radiation transport in 
hydrodynamical simulations \citep{Dessart2005} finds that damping of 
lateral velocity fluctuations isolates azimuthal zones, leading to azimuthal 
incoherence down to the grid scale. This might indicate that the wind clumps 
might be rather somewhat elongated in radial direction \citep[see 
also][]{Gomez2003}.  

In either case, the continuum opacity of the wind consisting of such structures 
is anisotropic. Isotropic opacity is only adequate if the wind clumps are 
spherical.}

\begin{figure}[t]
\centering
\epsfxsize=0.99\columnwidth
\mbox{\epsffile{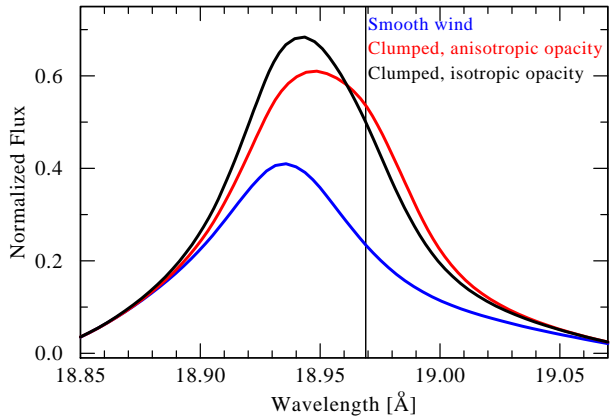}}
\caption{Three model line profiles of O\,{\sc vii} in the X-ray spectrum 
of a O9II  star.  Except of clumping properties, all other model 
parameters are the same, $\log{\dot{M}}=-6.4$\,[\myr], 
$v_\infty=2000$\,km\,s$^{-1}$, $\beta=0.8$. 
The onset of X-ray emission is at $1.2\,R_\ast$. The blue line shows the 
line emerging from a smooth wind. The red  line is a clumped wind model with
$L=1.3$ ($\approx 2\times 10^4$ clumps between $1.2\,R_\ast$ and  
$100\,R_\ast$) and an anisotropic opacity (the clumps are flat in 
radial direction). The black line is for the same parameters, 
but now for isotropic opacity (e.g.\ the clumps have spherical shape).  
The vertical line denotes the rest frame line frequency. 
The model profiles are convolved with the instrumental profile of the 
HETGS MEG spectrograph on board of the {\em Chandra} X-ray telescope.  
}
\label{fig:o7}
\end{figure}

The general  solution of radiative transfer for the case of 
continuum non-gray 
opacity in an expanding inhomogeneous stellar wind was presented by 
\citet{feld2003}. The optically thin 
and thick clumping as well as  bridging cases were included in the theory.  The 
formalism was developed in the context of X-ray lines emitted by an optically 
thin hot plasma that is attenuated in a fragmented cool stellar wind by strong 
continuum opacity. An analytic description for the effective isotropic and 
anisotropic opacity  was found. It was pointed out that in case of isotropic 
opacity (e.g.\ spherical clumps) the line {\em profiles} are 
identical to those emerging from a smooth wind. On the other hand, in case of  
anisotropic opacity, the line profiles are characteristically different 
\citep{osk2006}. The statistical solutions were verified by  2.5-D\,Monte-Carlo 
simulations \citep{osk2004} that allow non-constant and non-monotonic 
distributions of mass-absorption coefficients and filling factors. 
Example model X-ray lines are shown in Fig.\,\ref{fig:o7}.
{\changed As can be seen 
from this figure, the lines computed with clumped wind models are stronger 
compared to those computed with the smooth wind model for the same 
mass-loss rate. Also, the anisotropic wind opacity leads to a more symmetric 
and less blues-shifted line profile. }  

Assume that the flow of clumps that constitutes the wind obeys the
equation of continuity. The number of clumps per unit volume is
$n(r)$$\equiv$$n_0\,v(R)^{-1}R^{-2}$, where $n_0$\,[s$^{-1}$] is a
constant. The effective opacity of the clumped wind, $\kappa_{\rm eff}$, 
is the product of the clump number density $n(r)$, clump cross-section
$\sigma_{\rm C}$, and the probability of an X-ray photon getting
absorbed when it encounters a clump ${\cal P}=1-\exp{(-\tau_{\rm 
C}(\nu))}$, where $\tau_{\rm C}(\nu)$ is the optical depth of the
clump. The latter can be expressed as 
\begin{equation}
\tau_{\rm C}(\nu)=\kappa_{\nu} D \rho l\Rstar,
\label{eq:ctau}
\end{equation}
where  $r=R/\Rstar$ and $l$\, is the geometrical size of the clump 
expressed in $\Rstar$, and $\kappa_{\nu}$ [cm$^2$\,g$^{-1}$] is the mass 
absorption coefficient {\changed determining the 
continuum opacity} \citep{feld2003,osk-a2011}. 
Due to the strong wavelength dependence of $\kappa_\nu$,  
a clump may be optically thick at longer wavelengths but thin 
at shorter ones.
Evaluating the wind optical depth as an integral over effective
opacity along the line-of-sight $z$ gives the optical depth:
\begin{equation}
\tau_{\rm w} = n_0 \int_{z_{\rm em}}^\infty \frac{\sigma_{\rm C}}{v(r)\,r^2}
(1-{\rm e}^{-\tau_{\rm C}(\nu)})~{\rm d}z~~.
\label{eq:tcw}
\end{equation}
{\changed 
Note that this integral starts from the $z$ coordinate of the X-ray
emitter. The optical cross-section of spherical clumps (balls) is
isotropic: $\sigma_{\rm C} \propto r^2$. For clumps in the form of 
shell-fragments (pancakes), the cross section depends on the projection
angle as $\sigma_{\rm C} \propto |\mu|r^2$, where $\mu$ is the direction
cosine. For the latter case, the integral Eq.\,\ref{eq:tcw} has a
special property, because ${\rm d}z = {\rm d}r/\mu$. Hence, for X-ray
line  emitters in the front hemisphere which are located at some given
radius,  the $\mu$ dependence cancels out, i.e.\ all blue-shifted
frequencies  encounter the same optical depth and thus the same wind
absorption.  Only for the emitters in the back hemisphere, which create the
red line wing,  the optical depth increases with increasing red-shift
\citep{osk2004}.  

In case when all clumps are optically thick, the emergent flux in 
the line is determined by the radiation leaking between these opaque clumps, 
and depends only on the clump shape and their geometrical distribution 
\citep{feld2003,osk2004}. On the other hand, when all clumps are optically 
thin, the radiative transfer is not affected by the geometrical distribution of 
clumps and the line shape is the same as in case of smooth winds. It is 
important to remember, however, that the clump optical depth is a function of 
wavelength. Thus, it is most likely that neither of these two limiting cases 
(only optically thin or thick clumps) describe realistic winds. 

Thus a more gray opacity and more symmetric lines are expected from  
clumped winds compared to  smooth ones. This agrees with observations, 
namely nearly symmetric X-ray line profiles, and the similarity of their 
profiles in the soft and hard parts of spectrum 
\citep{wc2001,WC2007,Kramer2003}. An alternative explanation could be provided 
by reduced mass-loss rates \citep{Kramer2003, leu2013}.}

\citet{oc2006} and \citet{leu2013} used their porosity formalism  to 
describe X-ray lines emerging from inhomogeneous stellar winds. They 
found that a substantial reduction in wind absorption requires quite 
large porosity lengths, {\changed and claimed that such large  
lengths ($\sim 1R_\ast$) are unphysical. The basis for these claims is the  
outcome from first 2-D models of non-stationary winds that were not capable 
to produce laterally coupled structures in the winds \citep{Dessart2005}. 
However, these isothermal models are not capable to explain the X-ray emission 
from stellar winds either. Therefore it is not yet clear how qualitatively 
robust are their predictions on the size of the clumps. Moreover, the cool wind 
opacity has a strong wavelength dependence. Therefore, for harder radiation 
the clumps will remain optically thin even for very large porosity lengths, 
while 
for the softer radiation even small clumps will be optically thick. E.g.,\ 
in the wind of the prototypical O star $\zeta$\,Pup, clumps located at 
$2\,R_\ast$ and with geometrical size of $0.07\,R_\ast$ will be optically 
thick for the  radiation in the  Ne\,{\sc x} $\lambda 12.13$ line  
\citep{osk-a2011}. }

\citet{Herve2012} provided a careful comparison of the macroclumping and 
porosity formalisms and concluded that they are ``essentially equivalent 
to first order''. In a way this is not surprising, because in the porosity 
formalism the exact integrals \citep{feld2003,osk2006} are simplified using 
Taylor expansions. \citet{Herve2012} modeled the X-ray spectra of O-stars. 
They concluded that including porosity does not improve the line fits. On the 
other hand, they used the radial dependence of X-ray filling factor as a 
model parameter. They found that such radial dependence is 
required to provide a suitable fit to the observed X-ray spectrum of 
an O supergiant. \citet{leu2013} incorporated the porosity formalism in a 
standard 
X-ray spectra fitting software. Fitting observed X-ray spectra, 
they found that neither porosity nor radial dependent filling factors 
improve the line fits, and concluded that these effects are not important and 
can be neglected. However, {\changed all recent studies of X-ray spectra of O 
stars by means of sophisticated non-LTE models show that simple smooth wind 
models with constant filling factors are incapable to 
explain the multiwavelength spectroscopic observations. Either macroclumping 
or radially dependent X-ray filling factors  have to be included in the 
models to reproduce the observations adequately 
\citep{Herve2012,Shenar2015,Rauwa2015,Puebla2016}.}

Apart from the absorption of X-rays in clumped stellar winds, the problem 
of X-ray absorption in the interstellar medium also requires an adequate 
treatment of inhomogeneities. \citet{Wilms2000} provided a solution for 
this problem by including the self-shielding of the dust grains, and derived 
analytical  expressions for the optical depth of the grains, assuming that 
the grains can be approximated as spheres. 

\subsection{\em Line opacity}
\label{sec:macro}

\citet{osk2007} realized that since the optical depth in the UV resonance 
lines is high and the line photon mean free path is short, the wind 
inhomogeneities are likely to be optically thick at these wavelengths. This is 
in agreement with \citet{Prinja2010},  who found spectroscopic evidence of 
optically thick clumps in the winds of B supergiants by measuring the ratios 
between  the radial optical depths of the red and blue components of the 
Si\,{\sc iv} doublets. From atomic physics this ratio should be exactly two. 
However, in all observed stars, only values $< 2$ were found. This was 
interpreted as  a direct signature of optically thick clumping. 

\begin{figure*}[t]
\centering
\epsfxsize=0.8\columnwidth
\mbox{\epsffile{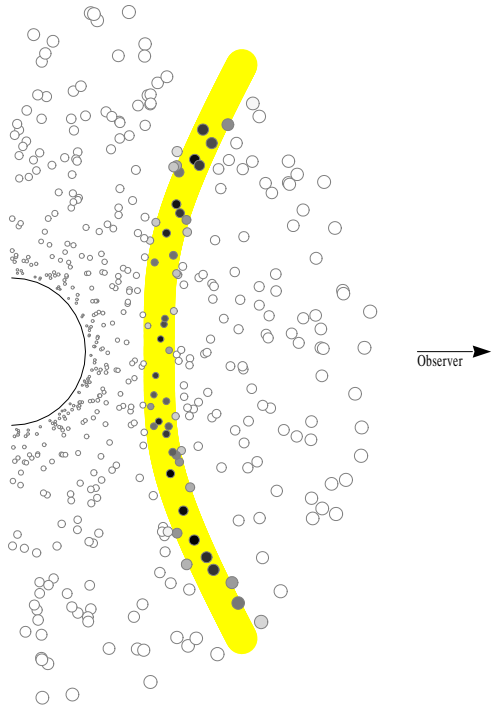}}
\caption{
Sketch of a clumped stellar wind. In a smooth wind, rays of a
given observer's frame frequency encounter line opacity only close to
the ``constant radial velocity surface'' (thick shaded line). In
a clumpy wind, assuming that the clumps move with the same velocity
law  as for the homogeneous wind, only those clumps interact
with the ray that lie close to the corresponding constant radial velocity 
surface (dark-shaded circles). All other clumps are transparent (open circles) 
if the continuum opacity is small, so the wind is porous with respect to line
transfer, even when the total volume is crowded with clumps. Adopted 
from \citet{osk2007}}
\label{fig:sketch}
\end{figure*}

To understand how optically thick clumping affects the radiative transfer in 
resonance lines, it is useful to consider the Sobolev approximation. According 
to this approximation,  only matter close to the constant radial velocity 
surface contributes to the line optical depths. In a clumped wind, this surface 
will be porous (Fig.\,\ref{fig:sketch}). Moreover, the opacity depends not only 
on the geometrical matter distribution, but also on the Sobolev length. For a 
smooth monotonic velocity, the Sobolev length is given by 
$v_{\rm D} ({\rm d}v/{\rm d}r)^{-1}$. {\changed In the original Sobolev 
approximation, $v_{\rm D}$ denotes the Doppler width by thermal broadening. 
Generalized to a clumpy medium, $v_{\rm D}$ can be considered as the dispersion 
of velocities inside an individual clump \citep{osk2007,Surlan2012}. Hence, 
the smaller the velocity dispersion within each
clump, the narrower is the constant radial velocity surface. }Consequently, a 
smaller number of clumps can contribute to the effective opacity, farther 
reducing it. In principle, the parameter $v_{\rm D}$ could be estimated from 
observations using spectroscopic 
techniques \citep[e.g.][]{Surlan2013,Simon2014,Sander2015}.

\citet{Owocki2008} considered the special situation of a strongly non-monotonic 
wind velocity, motivated by  dynamical simulations of the line-driven
instability \citep{Owocki1988, feld1997}. At present, these dynamic
models are limited to 1-D geometry, and it is not clear whether the same strong 
shocks would be seen in multi-dimensional models \citep{Gomez2003,Dessart2005}. 
Moreover, observations of X-ray variability in massive stars, so far, do not
confirm the 1-D model predictions on the temporal behavior of X-ray emission  --
no  X-ray variability attributable to  strong stochastic shocks is observed
\citep[e.g.][]{osk2001,naze2013}. \citet{Owocki2008} suggested that the strongly 
non-monotonic wind velocity seen in 1-D models should lead to a reduction of the
resonance line strength that is insensitive to spatial scales. This suggestion
was in detail examined using their 2-D models by \citet{Sund2010}, who indeed
found a reduction in the line strengths. However, the  3-D
models (see Section\,\ref{sec:pcl}) revealed that strongly non-monotonic
wind velocity has only moderate effect on the effective  opacity.
{\changed  Moreover, it seems of limited importance to discuss the 
porosity and non-monotonic velocity effects separately. Both are playing 
role in establishing the line opacity and are consistently included in the 
macroclumping formalism . }

To investigate the effects of macroclumping, an approximate treatment of the 
effective opacity was developed by \citet{osk2007}. A correction factor for 
macroclumping that can be easily included in a sophisticated non-LTE codes 
was derived.  {\changed This treatment introduces a next level of complexity 
compared to  microclumping, but is still quite  approximate. Nevertheless, 
the validity of this approach is confirmed by 3-D Monte Carlo models (see 
Section\,\ref{sec:pcl})}. 

In this macroclumping approximation, it is assumed that the clumps are 
statistically distributed, having an
average separation $L$ between their centers ($L(r)$ is expressed in 
$R_\ast$ and varies with radial location); hence, the volume filling factor is 
$f_V = l^3 /
L^3$, or
\begin{equation}
D = L^3 / l^3\ . 
\label{eq:dl}
\end{equation}

\noindent From Eq.\,(\ref{eq:dl}) it follows that
\begin{equation}
n(r) \equiv L^{-3} = D^{-1}\,l^{-3},  
\label{eq:nc}
\end{equation} 
where $n_{\rm C}$ denotes the number density of the
stochastically distributed clumps. 

{\changed Optical depth across a clump can be approximated as  
$\tau_{\rm C}=\kappa_{\rm C} D \rho l R_\ast$. This is similar to the 
Eq.\,(\ref{eq:ctau}), except that now $\kappa_{\rm C}$ is the non-LTE opacity
(see Eq.\,\ref{eq:kappa_D}). Recalling that $f_{\rm V}=D^{-1}$ yields}
\begin{equation}
\tau_{\rm C} = \kappa_{\rm f}\ D^{2/3}\ L  \ .
\label{eq:tau_C}
\end{equation}

The effective opacity $\kappa_{\rm eff}$ of the clumpy medium is obtained 
in analogy to the usual opacity from atomic absorbers:
\begin{equation}
\kappa_{\rm eff} = n \sigma_{\rm C}\ .
\label{eq:eff}
\end{equation}
Here $\sigma_{\rm C}$ is the effective cross section of a
clump.  ``Effective'' means that the geometrical cross
section, $l^2$, is multiplied by the fraction of photons that is
absorbed when crossing the clump \citep{feld2003},
\begin{equation}          
\sigma_{\rm C} = l^2\  \left( 1 - e^{-\tau_{\rm C}} \right)\ .
\label{eq:sig}
\end{equation}

Combining Eqs.\,(\ref{eq:nc}), (\ref{eq:eff}), and  (\ref{eq:sig}) yields 
\begin{equation}
\kappa_{\rm eff} = (D l)^{-1}\ 
\left( 1 - e^{-\tau_{\rm C}} \right)\ .
\label{eq:kapdl}
\end{equation}
Noticing that 
$ (D l)^{-1} = \kappa_{\rm f} / \tau_{\rm C}$, the scaling of the 
effective opacity with the opacity obtained in the microclumping 
approximation can be written as:
\begin{equation}
\kappa_{\rm eff} = 
\kappa_{\rm f}\ \frac{1 - e^{-\tau_{\rm C}}}{\tau_{\rm C}}\,\equiv \,
 \kappa_{\rm f}\ C_{\rm macro}\ .
\label{eq:kef} 
\end{equation}
The factor $C_{\rm macro}$ thus describes how macroclumping changes the
opacity, compared to the microclumping limit. Note that for optically
thin clumps ($\tau_C \ll 1$) the microclumping approximation 
($\kappa_{\rm eff} \approx \kappa_{\rm f}$)  is recovered.
For optically thick clumps ($\tau_C \msim 1$), however, the effective opacity 
is reduced by a factor $C_{\rm macro}$. {\changed Since a detailed radiative 
transfer calculations inside each clump is yet impossible, the same reduction 
factor is applied to the corresponding emissivities, i.e.\ the non-LTE source 
is unchanged.} 

Synthetic spectra were computed using this macroclumping formalism   
and compared to the observed spectra of an O-type supergiant \citep{osk2007}. 
It was shown that while the H$\alpha$ line is not affected by macroclumping, 
the  
P\,{\sc v} resonance doublet becomes significantly weaker and the mass-loss 
rate measured from it becomes higher when macroclumping is accounted 
for. Thus, it was demonstrated that including macroclumping in the 
spectral analysis resolves the problem of discordant mass-loss rates 
obtained using  $\rho$- and $\rho^2$-diagnostics. 

Importantly, \citet{Petrov2014} pointed out that for cooler B supergiants  
macroclumping should play a significant role in the formation of 
the H$\alpha$ line, altering H$\alpha$-based  mass-loss rates. 

To summarize, compared to microclumping which is mainly described by the 
parameter $D$ and its radial dependence, the macroclumping formalism requests 
at least one  additional parameter -- $L$. A  sensible choice for this 
parameter could be made from careful consideration of the observed spectrum.
\citet{Shenar2015} conducted a multiwavelength (X-ray to optical), non-LTE
spectroscopic analysis  of the bright eclipsing spectroscopic binary, 
$\delta$\,Ori\,A (O9.5II).  Motivated by hydrodynamic studies 
\citep{feld1997,Run2002}, they assumed that the clumping initiates at 
$r=1.1\,\Rstar$ and grows to its maximum contrast of $D = 10$ at $r\sim 
10\,\Rstar$. A radius dependent macroclumping parameter $L$ was used,
with $L=0.5\,\Rstar$ in the low wind regions.  The analysis confirmed the
strong effect of macroclumping on the resonance lines. The H$\alpha$ line, 
as well as the photospheric features, are hardly affected. The mass-loss 
rate derived from this multiwavelength spectral analysis was found to be in a 
good agreement with those theoretically predicted \citep{Vink2001}.

For comparison, a multi-wavelength (X-ray to optical) analysis, based on 
the non-LTE models, of the B0 Ia-supergiant $\epsilon$\,Ori did not include 
macroclumping.  In order to describe the resonance line of 
Si\,{\sc iv}, a microclumping parameter $f_\infty < 0.01$ had to be 
adopted \citep{Puebla2016}. The corresponding mass-loss rate is 
at least one order of magnitude lower than prescribed by the \citet{Vink2001} 
recipe. (As an alternative \citet{Puebla2016} discuss possible 
problems with their model wind ionization structure.)  

\section{Radiative transfer using realistic 3-D Monte-Carlo wind models}
\label{sec:pcl}

The statistical treatment of macroclumping briefly outlined in 
section~\ref{sec:macro} provides a first approximation for radiative 
transfer in clumped winds. While this approximation could be easy included 
in large codes, full 3-D models of clumped winds are required for 
in-depth studies. {\changed For the special case of line scattering, i.e.\ for 
resonance lines,} such model was developed by \citet{Surlan2012} and applied 
for the analysis of observed spectra by \citet{Surlan2013}. 

To study the basic effects of clumping on the resonance line formation
(both singlets and doublets) in the 3-D Monte-Carlo code, a core-halo model 
was adopted. Only the line opacity is taken into account. Pure scattering 
and Doppler broadening are considered, and complete redistribution is assumed.

To solve the radiative transfer throughout the clumped wind, first
a snapshot of the clump distribution is generated. Then, using the 
Monte Carlo approach the photons are followed along their paths. The density and
velocity of the wind can be arbitrarily defined in a 3-D space. The
calculations are carried out in the comoving frame following the
prescriptions by \citet{Hamann1980}.

The velocity field has  three arbitrary Cartesian components. For
simplicity, the velocity field is assumed radial and for the underlying
smooth wind the  standard $\beta$-velocity law is adopted. Non-monotonic
velocities are also incorporated  
\citep{Surlan2012}. The velocity inside the $i$-th clump is parametrized
as
\begin{equation}
\label{veldev}
v(r)=v_{\beta} (r^{\rm c}_{\rm i}) - m\,v_{\beta}(r) \,
\frac{r-r^{\rm c}_{\rm i}}{l_{i}},
\end{equation}
where $l_{i}$ is the radius of the $i$-th clump, $r^\mathrm{c}_{i}$ is the
absolute position of the center of the $i$-th clump,
$v_\beta(r^\mathrm{c}_{i})$ is the velocity determined according to
$\beta$-velocity law at the position $r^\mathrm{c}_{i}$, and the velocity
dispersion is $v_{\rm dis}(r)=m\,v_{\beta}(r)$, where $m$ ($ 0 < m\le 1$)
is the velocity deviation parameter (free parameter of the model).
Thus, Eq.\,(\ref{veldev}) introduces a negative velocity gradient inside
clumps, while the center of the clump moves according to a $\beta$-velocity 
law.

For the opacity, the parametrization of \citet{Hamann1980} is used.
Allowing for an arbitrary optical depth, clumps can be optically thick in
the cores of resonance lines, while they remain optically thin at
all other frequencies. For simplicity a spherical shape of the clumps is
assumed, with a radius that varies with the distance $r$ from the star,
$l=l(r)$. The density inside clumps is assumed to be
higher  by the clumping factor $D$ than the smooth wind density at same $r$.

While for the statistical treatment we assumed that the interclump medium is 
void (Section\,\ref{sec:osk07}), observations seems to indicate otherwise.
The presence of a tenuous interclump medium is supported by the
X-ray emission from O-type stars and by their strong lines of overionized  
ions, e.g.\ O{\sc vi} $\lambda\lambda 1032, 1038$\,\AA\  
\citep{co1979,Zs2008}. The interclump medium
is specified  in the 3-D Monte Carlo models by the 
interclump density parameter $d$ ($0\le d < 1$), which
is assumed to be depth independent. The interclump medium is rarefied by the 
factor $d$ compared to
the smooth wind density. For the case of dense clumps and non-void interclump
medium, the mass of the wind is distributed between clumps and interclump 
medium, and the clump filling factor becomes $f_{V}=(1-d)/(D-1)$, which may be 
depth dependent.

\begin{figure}[t]
\centering
\epsfxsize=0.99\columnwidth
\mbox{\epsffile{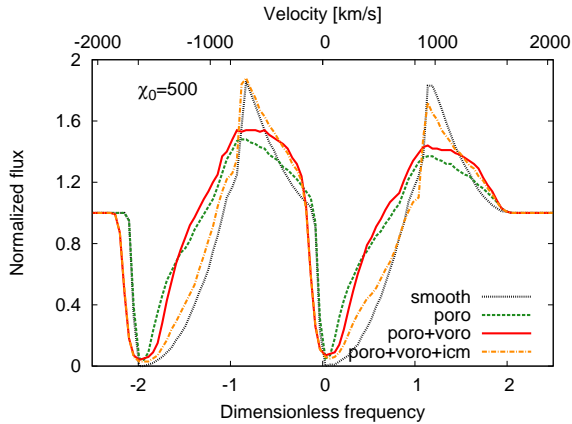}}
\caption{Effect of the macroclumping on a strong resonance doublet. The 
model accounts for a non-void interclump medium and for a velocity dispersion 
within the  clumps. Four model lines are shown, computed with  the same 
mass-loss rate. The  black dashed line shows a smooth wind model, the green 
dashed lines 
illustrates the effect of macroclumping. The solid red line shows the same 
model as the green line, but now including velocity dispersion within clumps. 
The orange dash-dotted line shows the full model that includes also the effect 
of not-void interclump medium.  
Adopted from Fig.\,10 in \citet[][]{Surlan2012}}
\label{fig:porl}
\end{figure}

The detailed study  by \citet{Surlan2012} showed that the different values for 
the parameters 
describing the clumping and the velocity field result in different 
strengths and shapes of the resonance lines (Fig.\,\ref{fig:porl}). 
The line profiles are sensitive to the spatial distribution of wind 
clumping. The density contrast, the clumping onset radius, and its 
radial distribution -- all these play a role for the line formation.  

Overall, the 3-D models  confirm that macroclumping
reduces the effective opacity in the resonance lines, and conclusively prove 
that in a realistic 3-D wind with density and velocity variations the P Cygni
profiles from resonance lines are different compared to smooth and stationary 
3-D winds.  The parameter study showed that the key model parameter affecting 
the effective opacity is the clump separation, $L$.

Besides these general results, the 3-D models of radiative transfer in
inhomogeneous stellar winds also brought new important insights that 
allow for a detail comparison with observations. E.g.,\ the presence of an 
absorption  dip near $v_\infty$ in a line
profile  was explained by the weakening of 
the macroclumping effects in outer regions of the stellar wind.  A new 
diagnostic 
method for the onset of wind clumping was suggested, using the absorption dip 
at the line center. The study also confirmed the importance of the interclump 
medium, by demonstrating that a non-void interclump medium is required to 
reproduce the saturated spectral lines simultaneously with non-saturated ones. 
Finally, the study has highlighted that in any realistic wind, non-monotonic
velocity is always connected with density inhomogeneity.  Therefore, their 
combined effects must be accounted for.   

To compare the 3-D model with observations and measure mass-loss rates and
other wind parameters, \citet{Surlan2013} analyzed the optical and
UV spectra of five O supergiant stars. This was done using a combination
of the  PoWR non-LTE stellar atmospheres and the Monte Carlo routine for the
transfer of radiation in resonance lines.

\begin{figure}[t]
\centering
\epsfxsize=0.99\columnwidth
\mbox{\epsffile{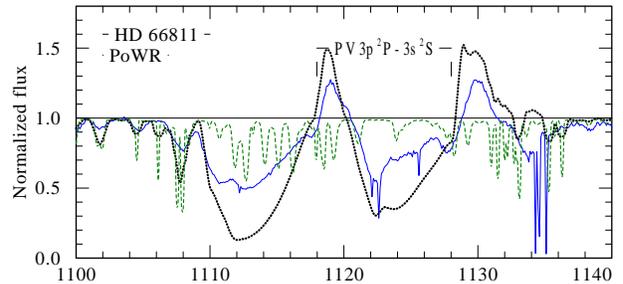}}
\caption{Comparison of observed and model spectra around  P\,{\sc v} 
resonance doublet in the 
O4I star HD\,66811 ($\zeta$ Pup). The thin solid-blue line is the observed 
spectrum. Dotted black is the PoWR model spectrum adopting 
$\dot{M}=2.5\times 10^{-6}$\,\myr\ {\changed and a depth dependent 
microclumping filling factor, which starts to deviate from the homogeneous 
wind ($D=1$) at about the sonic point (5\,km\,s$^{-1}$) and reaches 
$D=10$ at $v(r)=40$\,km\,s$^{-1}$. The dashed-green lines are from the same
model, but only accounting for the photospheric spectrum while the wind 
contribution is suppressed. Not accounting for macroclumping, the PoWR model 
predicts a much stronger P\,{\sc v} doublet than observed.}}
\label{fig:pvpowr}
\end{figure}

\begin{figure}[t]
\centering
\epsfxsize=0.99\columnwidth
\mbox{\epsffile{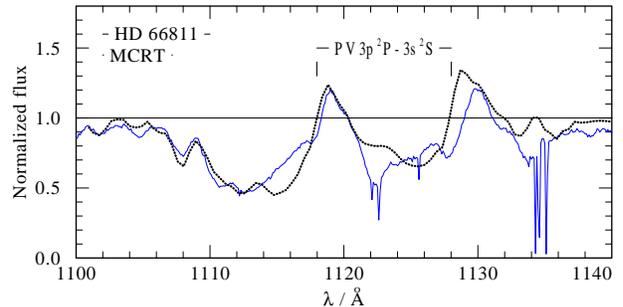}}
\caption{The same as in Fig.\,\ref{fig:pvpowr}, but now 
the dotted black line is computed with the  3-D Monte Carlo wind model, using 
only the PoWR photospheric spectrum as input. The adopted mass-loss rate is 
again $\dot{M}=2.5\times 10^{-6}$\,\myr. The line strength is significantly 
reduced compared to  Fig.\,\ref{fig:pvpowr} despite the same adopted 
$\dot{M}$. See model details in \citet{Surlan2013}.}
\label{fig:pvmc}
\end{figure}

As the first step, the PoWR models were fit to the observed optical spectra. 
From fitting the H$\alpha$ line, mass-loss rate and the microclumping 
parameters were obtained and then fixed. The abundances, ionization 
stratification, and underlying photospheric spectra were also adopted from 
the PoWR models. Confirming previous studies, the UV resonance lines 
(especially P\,{\sc v}) were found to be too weak for the H$\alpha$ based 
mass-loss rate (Fig.\,\ref{fig:pvpowr}). Therefore, for these 
lines the 3-D Monte Carlo code was applied. The fixed, H$\alpha$ based 
mass-loss rate was not changed during fitting, but the clumping parameters 
(density and  velocity 
fields) could be adjusted (Fig.\,\ref{fig:pvmc}).  
 
It turned out that the H$\alpha$ and the P\,{\sc v} lines can be fitted with 
the same mass-loss rate. Thus macroclumping resolves the discordance 
in mass-loss 
diagnostics as being due to the deficiencies in the treatment of clumping 
in standard modeling. The study also showed that the mass-loss rates obtained 
using macroclumping  are only 1.5 to 2.5 times lower than predicted by 
the standard mass-loss recipe \citep{Vink2001}. The number of clumps required 
to obtain good fits, $\sim 10^4$ clumps up to 100\,$R_\ast$ compares well with 
the numbers observationally deduced from optical line  variability studies 
and from  X-ray observations of HMXBs \citep{Eversberg1998, Furst2010}.

\section{What are the true empirical mass-loss rates of massive stars?}
\label{sec:true}

{\changed
The problem of realistic stellar mass-loss rates remains in the focus of
massive star astrophysics.  The concept of macroclumping, implemented so
far as a statistical approach and as 3-D Monte-Carlo modeling for
resonance lines, improved  the empirical mass-loss diagnostics.
Importantly, applications of these models revealed that the
empirical mass-loss rates for O-stars that have been determined in the
last years on the basis of the microclumping approach must not be
drastically revised. The mass-loss rates of massive stars broadly 
agree with theoretical predictions \citep[such as ][]{Vink2001}. 
As a fortunate consequence, the mass-loss prescriptions used in  
the established evolutionary models for  massive stars
\citep[e.g.][]{Brott2011,Georg2012} remain valid, at least roughly. 

Since strong wind clumping modifies the effective opacity, it might 
also affect the radiative driving mechanism  
\citep[e.g.][]{Mui2012,Sund2014}. The usual models that predict O-star mass
loss rates \citep{Vink2001} do not account for this.  Due to the
complexity  of the problem and the likely connections between wind
clumping and  sub-photospheric structures \citep{Jiang2015}, no
self-consistent models  of radiatively driven inhomogeneous stellar
winds exist yet. The value of the mass-loss rate is determined by the
physical conditions at the critical point at the base of the wind.
Since the empirical mass-loss rates accounting for macroclumping, as
reported above, are consistent within a factor of 1--3 with those 
predicted by the models from \citet{Vink2001}, one may speculate that 
clumping is not yet strongly developed in these low-velocity layers. 

\citet{Vink2012} has shown that the mass-loss rates provided by current 
theoretical models for the Of/WNh-type stars are of the right order of 
magnitude. In evolved Wolf-Rayet stars the mass-loss rates are large
and mainly diagnosed from their emission-line spectra 
\citep[e.g.][]{Hamann2006,Sander2012, Hainich2014}. Similar to the
O-stars, macroclumping may be required to achieve full accordance of 
mass-loss rate estimates using the UV resonance lines 
\citep{Kub2015}.

In this paper we did not cover the B and O-type stars of lower
luminosities. The mass-loss rates estimated for these stars, based on 
IR, optical, and UV diagnostics, are much lower than expected 
\citep[e.g.][]{Martins2005, Marcolino2009, Naj2011}, posing the
so-called  ``weak wind problem''.  Macroclumping alone cannot fully
resolve this problem. A possible explanation is that a significant
fraction of the wind in these stars is in a shock-heated phase and,
therefore,  can be detected only in X-rays 
\citep[e.g.][]{Cas1994,Drew1994,Hue2012,Lucy2012}.

Despite the success of the macroclumping model, one has to be aware of
its limitations, such as a number of additional free parameters,
and the approximate treatment of the physical conditions within  the
clumps. Nevertheless, macroclumping is a step forward in our quest for 
realistic descriptions of stellar wind, which would not have been 
possible without the deep insights of V.V.\,Sobolev and his school on the
physics of moving  stellar envelopes. 

\bigskip
The authors are indebted to an anonymous referee for the detailed and 
important comments and suggestions that significantly improved this paper.  } 
LO and BK acknowledge support from the DLR grant 50 OR 1508 and 
GA \^CR grant 16-01116S respectively. 
 

\end{document}